\documentclass[aps,prd,twocolumn,superscriptaddress,nofootinbib,showpacs,10pt,longbibliography]{revtex4-1} 
\usepackage[left=2.0cm,right=2.0cm,top=2.0cm,bottom=2.0cm,marginparwidth=17mm]{geometry}
\usepackage[utf8]{inputenc}
\usepackage[english]{babel}
\usepackage{amsmath}
\usepackage{amsfonts}
\usepackage{amssymb}
\usepackage{graphics}
\usepackage{graphicx}
\usepackage[usenames,dvipsnames,svgnames]{xcolor}  
\usepackage[linktocpage]{hyperref}   
\usepackage{todonotes}
\usepackage{ulem} 
\hypersetup{colorlinks=true,linkcolor=beamer@PRD, citecolor=beamer@PRD}
\newcommand*\doperator{\mathop{}\!\mathrm{\,d}}

\renewcommand\eqref[1]{\textcolor{beamer@PRD}{(}\ref{#1}\textcolor{beamer@PRD}{)}}

\definecolor{beamer@PRD}{RGB}{46,48,146}
\begin{document}
\title{An introductory review on resource theories of generalized nonclassical light}
\author{Sanjib Dey}\email{sanjibdey4@gmail.com}\email{dey@iisermohali.ac.in}
\affiliation{Department of Physical Sciences, Indian Institute of Science Education and Research Mohali, Sector 81, SAS Nagar, Manauli 140306, India}
\begin{abstract}
Quantum resource theory is perhaps the most revolutionary framework that quantum physics has ever experienced. It plays vigorous roles in unifying the quantification methods of a requisite quantum effect as wells as in identifying protocols that optimize its usefulness in a given application in areas ranging from quantum information to computation. Moreover, the resource theories have transmuted radical quantum phenomena like coherence, nonclassicality and entanglement from being just intriguing to being helpful in executing realistic thoughts. A general quantum resource theoretical framework relies on the method of categorization of all possible quantum states into two sets, namely, the free set and the resource set. Associated with the set of free states there is a number of free quantum operations emerging from the natural constraints attributed to the corresponding physical system. Then, the task of quantum resource theory is to discover possible aspects arising from the restricted set of operations as resources. Along with the rapid growth of various resource theories corresponding to standard harmonic oscillator quantum optical states, significant advancement has been expedited along the same direction for generalized quantum optical states. Generalized quantum optical framework strives to bring in several prosperous contemporary ideas including nonlinearity, $\mathcal{PT}$-symmetric non-Hermitian theories, $q$-deformed bosonic systems, etc., to accomplish similar but elevated objectives of the standard quantum optics and information theories. In this article, we review the developments of nonclassical resource theories of different generalized quantum optical states and their usefulness in the context of quantum information theories.
\end{abstract}
\maketitle
\tableofcontents
\section{\textbf{Introduction}} \label{sec1}
The value of a substance is a relative property that is dictated by its inadequacy, which means that the objects that are easily available in nature are not valuable. A resource theory is nothing but a systematic extension of this fundamental principle of economics. The basic idea of a resource theory is to classify the allowed (also called free) and prohibited set of actions for a given purpose, and then to figure out what can be achieved out of the allowed operations. Certain objects that are not possible to generate via the free operations are considered to be resources in the given framework. For example, consider a person is going out for a nearby day-trip. Certain things like nonperishable foods, sun-glass, money, etc., are allowed to carry. Suppose that the person does not own a car and, therefore, accompanying a personal car is prohibited in the given conjecture. Since carrying the money is allowed, the person is free to hire a cab or to catch a mode of public transport. Thus, a hired car or public transport is a resource of the day-trip. Likewise, fuel is the resource for an automobile driver and a cooking ingredient requiring refrigeration is a resource for a group of campers.

Quantum resource theory \cite{Chitambar_Gour_2019,Liu_Bu_Takagi_2019} is a generic framework to study the resources of the atomic and sub-atomic phenomena that are governed by quantum mechanical laws. Such frameworks systematically unify different aspects of a particular quantum phenomenon by removing the associated arbitrariness and ambiguities. Technically, the fundamental goal of a quantum resource theory is to establish a framework for quantifying a quantum notion via a restricted set of operations, while a full set of operations being implied to all the physically allowed processes in quantum mechanics. If one defines a permissible set of free quantum operations through which only some of the physically comprehensible states can be prepared, the complement of the free set is coined as the resource set. A state that neither belongs to the free set nor can be generated via the associated free quantum operations is called a resource state.

In recent days, quantum resource theory has become an immense area of interest to people working in quantum information theories. Currently, several quantum resource theories including the resource theories of coherence \cite{Baumgratz_Cramer_Plenio_2014,Winter_Yang_2016,Streltsov_etal_2017,Tan_etal_2017, Streltsov_Adesso_Plenio_2017_Review}, nonclassicality \cite{Asboth_Calsamiglia_Ritsch_2005,Yadin_etal_2018,Chitambar_Gour_2019,Kwon_etal_2019,Tan_Jeong_2019_Review,Ge_etal_2020, Ferrari_etal_2020}, quantum memory \cite{Rosset_Buscemi_Liang_2018}, nonlocality \cite{Wolfe_etal_2020,Rosset_Schmid_Buscemi_2020}, EPR steering \cite{Gallego_Aolita_2015}, quantum network \cite{Pirandola_2019}, imaginarity of quantum mechanics \cite{Wu_etal_2021}, quantum hypothesis testing \cite{Pirandola_etal_2019}, optimization of programmable quantum computers \cite{Banchi_etal_2020}, etc., are under intensive scrutiny. Whereas the same for quantum entanglement \cite{Horodecki_etal_2009,Eltschka_Siewert_2014,Killoran_Steinhoff_Plenio_2016, Pirandola_etal_2017, Contreras-Tejada_Palazuelos_de-Vicente_2019} has become the most acclaimed example. Entanglement is a resource for long-distance communication between two quantum labs with respect to the free operations being local quantum operations and classical communication (LOCC). When Alice and Bob are separated by a long distance, as far as our current technology is concerned, the only mode of communication is the classical channel, like a telephone. Since Alice and Bob each own a quantum lab, they are allowed to prepare different quantum states locally via quantum operations. However, since the communication channel is restricted to be classical, the local quantum states cannot be shared with each other to prepare a joint quantum state that is entangled. But, it is known that a pair of quantum states having opposite spins is inherently entangled even if they are located far apart and, thus, entanglement is a resource in this scheme.

While the resource theories of various quantum phenomena have experienced rapid progress in recent years \cite{Streltsov_Adesso_Plenio_2017_Review,Chitambar_Gour_2019,Tan_Jeong_2019_Review}, a profound advancement for the same for generalized nonclassical states has also been noticed \cite{Dey_2015,Dey_Hussin_2015,Dey_Hussin_2016,Dey_Fring_Hussin_2017,Zelaya_Dey_Hussin_2018,Dey_Nair_2020}. This review article seeks to summarize the nonclassical resource theories of the generalized nonlinear and $q$-deformed quantum optical states and their versatility in the context of quantum information and computation.

\section{\textbf{Resource theories of nonclassicality: Preliminaries}}\label{sec2}
Classical physics facilitates a complete description of an object by a point in the phase space, however, the uncertainty principle strongly resists such a point-like characterization of quantum mechanical particles. The Glauber coherent state being a minimum uncertainty state without any preferred axial direction in the optical phase space, its dynamics resemble most closely to a classical system \cite{Dey_Fring_2013_PRA}. This is the reason why the coherent state is usually agreed to be the most accurate classical state of light \cite{Glauber_1963,Sudarshan_1963}. Any state having a quantum-like behavior with respect to coherent states is called nonclassical. The degree of nonclassicality is determined by the amount of quantumness that the state possesses. In reality, nonclassical states fill most of the space of the Hilbert space, whereas the classical-like states occupy only a corner of it. Despite the fact that the coherent states have ample usage in modern physics, it is well-established that the nonclassical states manifest genuine supremacy over coherent states in many applications. Furthermore, often it turns out that strongly nonclassical states are more beneficial than the weakly ones. Therefore, the quantification of nonclassicality is treated as an important objective, which can be achieved within the resource theoretical framework with ease.

\subsection{The general scheme}\label{sec21}
In order to find resources of nonclassicality, the first task is to define a set of classical-like states, say $\mathbb{S}$, which belongs to the Hilbert space $\mathcal{H}$. Naturally, a nonclassical state resides in the complement set of $\mathbb{S}$. Let us also consider that associated with the set of classical state $\mathbb{S}$, there exists a set of permissible operations $\mathbb{O}$, which is a strict subset of the set of all possible allowed quantum operations. The permissible operations $\mathbb{O}$ are restricted in such a way that if $\mathbb{F}\in \mathbb{O}$ as well as $\psi\in\mathbb{S}$ then $\mathbb{F}(\psi)\in\mathbb{S}$, where $\psi$ represents an arbitrary quantum state. We are now ready to devise a resource theory for nonclassicality. We may propose that, for a particular resource theory, there exists a non-negative measure $M(\psi)$ of nonclassicality, which satisfies the following three fundamental properties:
\begin{eqnarray}
&& M(\psi)=0\quad \text{if} \quad \psi\in\mathbb{S}, \label{Property1}\\
&& M(\psi)\geq M(\mathbb{F}(\psi)) \quad \text{if} \quad \mathbb{F}\in\mathbb{O}, \label{Property2}\\
&& M\left(\textstyle\sum_i q_i\psi_i\right) \leq \textstyle\sum_i q_iM(\psi_i). \label{Property3}
\end{eqnarray}
The property \eqref{Property1} is obviously the first and foremost requirement that the measure $M(\psi)$ becomes positive if and only if the state $\psi$ turns out to be nonclassical. Property \eqref{Property2} implies that, given the set of operations $\mathbb{O}$, the measure $M(\psi)$ is a monotonic function of nonclassicality. It means that if $\mathbb{F}\in\mathbb{O}$, the nonclassicality measure $M(\psi)$ must decrease monotonically under the operation $\mathbb{F}$. This property, in turn, ensures that the nonclassical states can neither be generated from classical ones, nor the nonclassicality of the state can be enhanced via the set of operations $\mathbb{O}$. The only possible way to increase the nonclassicality is by replacing the state $\psi$ with another nonclassical state, say $\chi$, so that $M(\psi)<M(\chi)$. Therefore, the sources of nonclassicality can be characterized as supplementary resources which are needed to overcome the restrictions inherent to the classical states $\mathbb{S}$ and operations $\mathbb{O}$. Property \eqref{Property3} is a convexity property, which reinforces the concept that the nonclassicality can never be increased by mixing two states $\psi$ and $\chi$ statistically. It should be noted that a statistical mixture of type $q\psi+(1-q)\chi$ is intrinsically a classical procedure and, therefore, such a process can never be allowed to generate extra nonclassicality.

The framework discussed above is a general procedure to formulate a resource theory of nonclassicality. For further understanding of the mathematical treatment of the resource theory, one may refer, for instance \cite{Coecke_Fritz_Spekkens_2016}. It is worth mentioning that so far it has not been possible to unify all types of existing nonclassicalities and provide a single nonclassicality measure for all of them. But, one is compelled to find the resources of each type of nonclassicality separately by associating a corresponding measure $M(\psi)$ that fits the framework mentioned above. In the rest of the section, we shall discuss the resource theory for some types of nonclassicalities.

\subsection{Resource theory of quasi-probability distribution functions}\label{sec22}
Since the Glauber coherent state $|\alpha\rangle$ is a predominant example of a classical-like pure state, a classical statistical mixture of them will obviously remain classical. The corresponding density matrix is given by
\begin{equation}\label{Density1}
\rho_{\text{c}}=\int P_\text{c}(\alpha)~|\alpha\rangle\langle\alpha|\doperator^2\alpha,
\end{equation}
with $P_\text{c}(\alpha)$ being a probability density, such that $\int P_\text{c}(\alpha)\doperator^2\alpha =1$. A similar type of representation is also possible for general mixed quantum states. In fact, Glauber \cite{Glauber_1963} and Sudarshan \cite{Sudarshan_1963} showed that any quantum optical state can be expressed in the overcomplete coherent state basis and, thus, the density matrix of the corresponding states can be represented in the following form
\begin{equation}\label{Density2}
\rho=\int P(\alpha)~|\alpha\rangle\langle\alpha|\doperator^2\alpha,
\end{equation}
where $P(\alpha)$ is familiar as the Glauber-Sudarshan's $P$-function. Notice the structural similarity of \eqref{Density2} with \eqref{Density1}. The only dissimilarity lies in the interpretation of the two functions $P_\text{c}(\alpha)$ and $P(\alpha)$. Since $P(\alpha)$ is applicable to general quantum states, it is interpreted as a quasi-probability distribution function rather than the usual positive probability density. In other words, although $P(\alpha)$ is normalized, i.e. $\int P(\alpha)\doperator^2\alpha =1$, it may not be positive always. The reason is that when an arbitrary quantum optical state is expressed in the overcomplete coherent state basis, one obtains extra weight functions that ensure the resolution of identity. These extra weight functions are solely responsible for the negativity of $P(\alpha)$ for some quantum optical states. It becomes a positive probability density only when the corresponding state is classical. The states for which $P(\alpha)$ is negative are termed nonclassical. 

There are some other quasi-probability distribution functions dictating the nonclassical properties of quantum optical states; such as Husimi $Q$-function \cite{Husimi_1940}, Wigner distribution function \cite{Wigner_1932}, etc., which are discussed in most of the standard textbooks of quantum optics \cite{Scully_Zubairy_1997_Book,Dodonov_Manko_2003_Book,Gerry_Knight_2005_Book,Klauder_Sudarshan_2006_Book, Walls_Milburn_2007_Book,Agarwal_2013_Book}. In a recent work \cite{Tan_Choi_Jeong_2020}, the authors have introduced $s$-parameterized quasi-probabilities $P_\text{s}(\alpha)$ with which they have unified the notion of nonclassicality for all the quasi-probability distribution functions in the framework of resource theory. As per their formalism, the s-parameterized measure of nonclassicality is given by
\begin{equation}\label{QuasiMeasure}
M_\text{s}(\rho)=\frac{1}{2}\left(\int |P_\text{s}(\alpha)|\doperator^2\alpha -1\right), \quad s\in [-1,1],
\end{equation}
whose value is entirely controlled by the negativity of the associated quasi-probability distribution functions. Here, $s=1,0,-1$ correspond to the cases of the $P$-function, Wigner function and the $Q$-function, respectively. It was shown that with the increase of the parameter $s$, $M_\text{s}(\rho)$ \eqref{QuasiMeasure} becomes an increasingly stronger measure of  nonclassicality. Thus, it clearly indicates that the negativity of the $P$-function is the strongest and the same for the $Q$-function is the weakest measure of nonclassicality among the three quasi-probability distribution functions.

A similar type of resource theory, albeit, from a different angle has been explored in \cite{Tan_etal_2017,Yadin_etal_2018}, where the set of free quantum operations $\mathbb{O}$ has been restricted to the operations that allow forward feeding of measurement outcomes and linear optical operations. Thus, such a resource theory is more useful to the scenarios where the measurement outcomes are utilized to determine the degree of nonclassicality. It was also shown that a nonclassicality measure of such type of theories obeys the three necessary properties \eqref{Property1}, \eqref{Property2} and \eqref{Property3} as discussed in the earlier subsection. For further details on nonclassical resource theories of quasi-probability distributions, we refer the readers to a recent review article \cite{Tan_Jeong_2019_Review}. 

\subsection{Resource theory of quadrature squeezing}\label{sec23}
In quantum optics, quadratures $\hat{x}$ and $\hat{p}$ are the field amplitudes of the quantized electromagnetic field, whose expressions are incidentally similar to the dimensionless position and momentum operators of the linear harmonic oscillator
\begin{equation}
\hat{x}=\frac{\hat{a}+\hat{a}^\dagger}{2}, \quad \hat{p}=\frac{\hat{a}-\hat{a}^\dagger}{2i}.
\end{equation}
Quadrature squeezing refers to a situation when the variance of at least one of the quadratures is less than the right-hand side of the type of uncertainty relation that the state obeys. Further interesting phenomena of quadrature squeezing in the cases of Heisenberg's and Robertson's uncertainty relations can be found in \cite{Dey_Fring_Hussin_2018_Review}.

While the resource theory of the quasi-probability distribution function as discussed in the previous subsection may be interesting and helpful, often it turns out that the computation of the distribution function corresponding to some states becomes laborious. In this subsection, we shall explore a relatively simpler but equivalent resource. Before delving into the details, let us start with an example of a Gaussian wave-packet, whose position and momentum quadrature variances, $\sigma_x$ and $\sigma_p$, are not equal to each other. Given that the quadratures are statistically uncorrelated, the $P$-function of the wave-packet can be written as \cite{Mollow_Glauber_1967}
\begin{equation}\label{PG}
P_{\text{G}} (\alpha)=\frac{1}{\mathcal{N}}\left(\frac{(\text{Re}\alpha-A)^2}{\sigma_x-1/2}+\frac{(\text{Im}\alpha-B)^2}{\sigma_p-1/2}\right).
\end{equation}
Here $\mathcal{N}$ represents the normalization constant, and $A,B$ denote the central position of the distribution in the real and imaginary axes of the complex $\alpha$-plane. It is obvious that the function \eqref{PG} cannot exist as a normalizable distribution if the variances of any of the two quadratures become less than $1/2$. Therefore, if at least one of the quadrature variances of a quantum state becomes less than $1/2$, the state does not correspond to a Gaussian distribution and, therefore, it is nonclassical. The statement not only applies to the Gaussian state, but to any arbitrary quantum state. In fact, one can write a general relation between the $x$-quadrature variance and the $P$-function of an arbitrary state \cite{Mollow_Glauber_1967} as
\begin{equation}
\sigma_x=\frac{1}{2}\int P(\alpha)\left[1+\left(\alpha^\ast+\alpha-\langle\hat{a}^\dagger+\hat{a}\rangle\right)^2\right]\doperator^2 \alpha.
\end{equation}
The above relation tells that the $P$-function can be a positive probability density provided that $\sigma_x>1/2$, otherwise $P(\alpha)$ will always be negative. A similar relation can also be derived for the $\hat{p}$-quadrature variance, which is not shown here. However, what we realize is that if either $\sigma_x<1/2$ or $\sigma_p<1/2$ holds for a quantum state, the corresponding $P$-function fails to be positive and the associated state will be nonclassical. Therefore, it is no longer essential to study the $P$-function and its corresponding resources if one gains the knowledge of quadrature squeezing. The equivalences of quadrature squeezing with the other two distribution functions, namely the $Q$-function and the Wigner function are unfamiliar. However, as we discussed in the previous subsection, the $P$-function is the strongest measure of nonclassicality and, therefore, as far as the resource is concerned, the study of Wigner and $Q$-functions are no longer essential. But, an analysis of quadrature squeezing will suffice.

Let us now discuss a formulation of the resource theory of quadrature squeezing \cite{Gehrke_Sperling_Vogel_2012, Vogel_Sperling_2014,Sperling_Vogel_2015}. Let us consider an arbitrary but fixed operator $\hat{g}$ associated with an experimental setup, with $\hat{g}\equiv\hat{g}(\hat{a},\hat{a}^\dagger)$ being an operator-valued function of the ladder operators. Let us also propose that a quantum state is said to be nonclassical if there exists an observable $\hat{g}^\dagger \hat{g}$ for which
\begin{equation}\label{NegativityS}
\langle:\hat{g}^\dagger \hat{g}:\rangle <0,
\end{equation}
where $:\cdot \cdot:$ denotes the normal ordering of the operators. Then, the Hermitian operators, $:\hat{g}^\dagger \hat{g}:$ and $\hat{g}^\dagger \hat{g}$, are observables of the corresponding setup.  Clearly, the second observable is positive semi-definite, but the spectrum of the first one may be negative. The negativity of the difference of the corresponding expectation values can be measure with $\delta$, such that 
\begin{equation}\label{delta}
\delta=\langle:\hat{g}^\dagger \hat{g}:\rangle-\langle\hat{g}^\dagger \hat{g}\rangle,
\end{equation}
which can be derived easily by following the standard methods of normal ordering. However, since $\langle\hat{g}^\dagger\hat{g}\rangle\geq 0$, the following relation is guaranteed
\begin{equation}\label{delta1}
\delta\leq\langle:\hat{g}^\dagger \hat{g}:\rangle < 0.
\end{equation}
Therefore, the relative nonclassicality of a quantum state for a designated measurement setup can be defined by a non-negative measure $M$ as follows
\begin{equation}\label{MeasureN}
M=\left\{ \begin{array}{ll}
\langle:\hat{g}^\dagger \hat{g}:\rangle /\delta\quad & \mbox{if~} \langle:\hat{g}^\dagger \hat{g}:\rangle<0 \\
0 & \mbox{otherwise} ~.\end{array}\right. 
\end{equation}
In other words, the measure $M$ defined in \eqref{MeasureN} quantifies the negativity of the LHS of \eqref{NegativityS} relative to the lower bound determined by the relation \eqref{delta1}. The relation \eqref{MeasureN} also suggests that if the negativity of $\langle:\hat{g}^\dagger \hat{g}:\rangle$ approaches the lower bound, the corresponding quantum state exhibits a maximum non-negativity of the relative measure $M$, i.e., $M=1$. Hence, the maximal nonclassicality is achieved when $\delta= \langle:\hat{g}^\dagger \hat{g}:\rangle$, i.e., $\langle\hat{g}^\dagger \hat{g}\rangle=0$.

In order to apply this formalism to find resources of quadrature squeezing, let us consider a perfect homodyne detection setup. The principle of homodyne detection technique relies on the measurement of the probability distribution of the phase-induced $\hat{x}$-quadrature operator
\begin{equation}
\hat{x}_\phi=\frac{1}{2}\left(\hat{a}e^{i\phi}+\hat{a}^\dagger e^{-i\phi}\right),
\end{equation}
of a given radiation mode \cite{Mandel_Wolf_1995_Book,Vogel_Welsch_2006_Book}. One can choose the operator $\hat{g}$ as the standard deviation of the quadrature, i.e., $\hat{g}\equiv\Delta\hat{x}_\phi=\hat{x}_\phi-\langle\hat{x}_\phi\rangle$, so that the quadrature squeezing effect can be quantified with the measure given in \eqref{MeasureN}. The value of $\delta$ follows from the fact that the variance of the quadrature differs from its normal-ordered value by that of the vacuum state quadrature, i.e., $\delta=\langle:(\Delta \hat{x}_\phi)^2:\rangle-\langle(\Delta \hat{x}_\phi)^2\rangle=-\langle(\Delta \hat{x}_\phi)^2\rangle_\text{vac}$. As discussed earlier, the maximum robustness of the measure is obtained when $\langle(\Delta \hat{x}_\phi)^2\rangle$ vanishes. However, in reality, it is hardly possible to face a circumstance when the value of $\langle(\Delta \hat{x}_\phi)^2\rangle$ is exactly zero. Rather, a realistic situation should refer to a scenario when the value is very small, say $\langle(\Delta \hat{x}_\phi)^2\rangle_\text{min}$. In short, when $\langle(\Delta \hat{x}_\phi)^2\rangle=\langle(\Delta \hat{x}_\phi)^2\rangle_\text{min}$, we obtain $M_\text{qs}=M^\text{max}_\text{qs}$, and, therefore, the maximum robustness in the measure can be expressed as
\begin{equation}\label{QSMeasure}
M^\text{max}_\text{qs}=1-\frac{\langle(\Delta \hat{x}_\phi)^2\rangle_\text{min}}{\langle(\Delta \hat{x}_\phi)^2\rangle_\text{vac}}.
\end{equation}
Thus, it is obvious that the state will become maximally nonclassical when the quadrature variance of the corresponding state $\langle(\Delta \hat{x}_\phi)^2\rangle$ becomes minimum with respect to the same for the vacuum state. Therefore, one may conclude that the maximal quadrature squeezing implies maximal nonclassicality. 

\subsection{Resource theory of sub-Poissonian photon statistics}\label{sec24}
The phenomenon of the photon number squeezing is characterized by the nature of the sub-Poissonian photon distribution, where the photon number distribution of a state is narrower than the average number of photons, i.e., $\langle(\Delta \hat{n})^2\rangle <\langle \hat{n}\rangle$, with $\hat{n}=\hat{a}^\dagger\hat{a}$ being the number operator of the radiation field. The optical noise of a sub-Poissonian light can go well-below the classical shot-noise limit (the lowest detectable noise level with the classical light) and, thus, sub-Poissonian light often turns out to be extremely useful, especially, for the purpose of quantum information processing. Experimental detection of sub-Poissonian light is carried out with photodetectors and the resulting beam of photoelectrons also show sub-Poissonian nature \cite{Short_Mandel_1983,Teich_Saleh_1985}.

In the framework of resource theory, the nonclassicality of a sub-Poissonian light can be characterized easily by associating the operator $\hat{g}$ with the standard deviation of the photon number
\begin{equation}
\hat{g}\equiv\Delta\hat{n}=\hat{n}-\langle\hat{n}\rangle.
\end{equation}
By following a similar logic of quadrature squeezing, we can write
\begin{equation}
\delta=\langle:(\Delta \hat{n})^2:\rangle-\langle(\Delta \hat{n})^2\rangle=-\langle(\Delta \hat{n})^2\rangle_\text{Pois.}=\langle\hat{n}\rangle.
\end{equation} 
The resulting nonclassicality measure reads as follows
\begin{equation}\label{NSMeasure}
M_\text{ns}=1-\frac{\langle(\Delta \hat{n})^2\rangle}{\langle(\hat{n})\rangle}=-\mathcal{Q},
\end{equation}
where $\mathcal{Q}$ denotes the Mandel parameter. The result is unambiguous in the sense that the nonclassical measure $M_\text{ns}$ attains a maximal value of $1$ when the photon number variance $\langle(\Delta \hat{n})^2\rangle$ vanishes, i.e., when the number squeezing is maximized. One of the interesting notions of the resource \eqref{MeasureN} utilized in this and the previous subsection is that the measures are directly related to the mean values of the observables, which can be measured experimentally. Thus, it becomes easier for an experimenter to detect robust nonclassical states through an analysis of quadrature and photon number squeezing.

\section{\textbf{Generalizations of quantum optics}}\label{sec3}
Before moving to the discussion of generalizations of nonclassical states, let us highlight the basic notions of non-Hermitian theories which will be utilized for the generalizations.

\subsection{Role of non-Hermitian theories}\label{sec31}
Observables in quantum mechanics demand Hermiticity of the corresponding operators, which assures reality of spectrum and unitary time-evolution. However, quantum physics has evolved a lot over the centuries and standing at the 21st century of physics, it is evident that Hermitian operators are only a subset of physical observables. The journey of non-Hermitian operators as observables started almost at the time of the inception of quantum physics. In the year 1929, von Neumann and Wigner \cite{von-Neumann_Wigner_1929} indicated the possibility that the non-Hermitian Hamiltonians can acquire real positive eigenspectrum and discrete eigenstates. Later, such systems went through a more intense investigation, and, nowadays, the notion of the so-called bound states in the continuum (BICs) is well-realized via different examples \cite{Friedrich_Wintgen_1985} along with their bi-orthonormal eigenfunctions \cite{Persson_Gorin_Rotter_1996}. Towards the end of the last century and at the beginning of this century, we have noticed a rapid growth in this direction, particularly in the $\mathcal{PT}$-symmetric theories \cite{Bender_Boettcher_1998} and the formulation of pseudo-Hermitian Hamiltonians \cite{Mostafazadeh_2002}. Increased interest in these two theories has been dispersed in almost every branch of physics, likewise in quantum optics and information theories with no exception. While the role of non-Hermiticity in optical waveguide theories has been crucial, almost inevitably it appears in the process of the generalization of quantum optical models also. This review being focused on generalized quantum optics, we shall mainly discuss the role of non-Hermiticity in such theories. Before that, let us briefly summarize the basic notions of the two main directions of non-Hermitian theories in the rest of this subsection.

\subsubsection{$\mathcal{PT}$-symmetry}\label{sec311}
It was Wigner \cite{Wigner_1960} who pointed out that any operator which is invariant under an anti-linear involution whose eigenfunctions also respect the symmetry possesses real eigenvalues. Bender and Boettcher, in a seminal paper \cite{Bender_Boettcher_1998}, have shown that the Hamiltonians which are symmetric under the combined operation of parity $\mathcal{P}$ and time-reversal $\mathcal{T}$, i.e., which are invariant under an anti-linear operation, like $x\rightarrow -x,~p\rightarrow p,~i\rightarrow -i$, admit real spectrum. The result is clearly in agreement with Wigner's theorem \cite{Wigner_1960} without any loss of generality. Note that there exist subtle issues for the non-existence of the real spectrum even when the non-Hermitian Hamiltonian is $\mathcal{PT}$-symmetric, i.e., $[H,\mathcal{PT}]=0$. Since $\mathcal{PT}$ is an anti-linear operator, it is not guaranteed that all of its eigenstates will be the eigenstates of the Hamiltonian. The case when the wavefunctions are simultaneous eigenstates of the Hamiltonian and the $\mathcal{PT}$-operator, is referred to as $\mathcal{PT}$ unbroken scenario and the eigenvalues are always real. Broken $\mathcal{PT}$ cases (i.e., when $[H,\mathcal{PT}]=0$ but $\mathcal{PT}\psi'\neq \psi'$) may exhibit conjugate pairs of complex eigenvalues and the reality of the spectrum is not guaranteed. Thus, the proper way to interpret the theory is that the non-Hermitian Hamiltonians with unbroken $\mathcal{PT}$-symmetry always possess real eigenspectrum \cite{Bender_2007_Review}. 

Thus, a non-Hermitian generalization of the standard quantum theory demands a replacement of the Hermitian adjoint ($H=H^\dagger$) with the $\mathcal{PT}$-adjoint ($H=H^{\mathcal{PT}}$), such that the inner product $\langle\psi'|\phi'\rangle^{\mathcal{PT}}$ of the corresponding Hilbert space takes the form
\begin{equation}
\int\left[\psi'(x)\right]^{\mathcal{PT}}\phi'(x)\doperator x = \int\left[\psi'(-x)\right]^\ast\phi'(x)\doperator x.
\end{equation}
Due to the fact that $\mathcal{PT}$ is an anti-linear operator, wevafunctions may not be simultaneous eigenfunctions of $H$ and $\mathcal{PT}$, which implies that the norm may not be positive definite always. However, the problem is resolved by introducing a charge conjugation operator $\mathcal{C}$ and replacing the $\mathcal{PT}$-norm with the $\mathcal{CPT}$-norm \cite{Bender_Brody_Jones_2002}. Since then it is widely accepted that within the regime of unbroken $\mathcal{PT}$-symmetry a $\mathcal{CPT}$-symmetric non-Hermitian Hamiltonian describes a complete self-consistent physical system with real eigenvalues and unitary time-evolution. 

Rigorous developments of $\mathcal{PT}$-symmetric theories have been carried out over the years in various mathematical \cite{Dorey_Dunning_Tateo_2001,Shin_2001,Weigert_2003,Dey_Fring_Gouba_2012} and physical contexts \cite{Dey_Raj_Goyal_2019}. Plenty of experimental confirmation of $\mathcal{PT}$-symmetric effects as well as their applications in various branches of physics have enriched the field. 
Early experiments have revealed materials with controllable loss and gain by drawing an analogy between the stationary Schr\"odinger equation and the Helmholtz equation describing monochromatic linearly polarized light \cite{Guo_etal_2009,Ruter_etal_2010,Regensburger_etal_2012}. This has given rise to meta-materials \cite{Feng_etal_2013} exhibiting the phenomena of unidirectional invisibility \cite{Lin_etal_2011} and coherent perfect absorption \cite{Chong_etal_2010}. $PT$-symmetry has also been applied in microwave cavities physics \cite{Bittner_etal_2012}, nuclear magnetic resonant quantum systems \cite{Zheng_Hao_Long_2013}, finding topological edge states \cite{Xiao_etal_2017,Mittal_Raj_Dey_etal_2021}, stimulating the superconductivity \cite{Chtchelkatchev_etal_2012}, etc. It is hardly possible to provide a complete list of articles devoted to the development of the theory, however, for more information and references, one may follow some of the recent review articles \cite{Bender_2007_Review,El-Ganainy_etal_2018,Dey_Fring_Hussin_2018_Review} and books on the filed \cite{Moiseyev_2011_Book,Bagarello_etal_2015_Book, Bender_2018_Book}.  

\subsubsection{Pseudo-Hermiticity}\label{sec312}
A second approach to studying non-Hermitian physics is popular as pseudo-Hermiticity \cite{Mostafazadeh_2002}. The initial idea came from the work of Dirac and Pauli \cite{Pauli_1943} in 1943, which was followed up by Sudarshan \cite{Sudarshan_1961}, Lee and Wick \cite{Lee_Wick_1969} to overcome the renormalization induced negative norm states during the process of quantizing the electromagnetic and other fields. It was even before the appearance of $\mathcal{PT}$-symmetry and pseudo-Hermiticity in the literature when people tried to provide a consistent quantum mechanical framework of non-Hermitian Hamiltonians using the concept of quasi-Hermiticity  \cite{Dieudonne_1961,Scholtz_Geyer_Hahne_1992}. The concept was resurfaced by Mostafazadeh in a more popular form as pseudo-Hermiticity \cite{Mostafazadeh_2002}, where a non-Hermitian Hamiltonian $H_{\text{nh}}$ and a Hermitian Hamiltonian $H_\text{h}$ are considered to be related by a similarity transformation
\begin{eqnarray}\label{Pseudo_Hermiticity}
&& H_\text{h}=\eta H_\text{nh}\eta^{-1}=H_\text{h}^\dagger=\eta^{-1}H_\text{nh}^\dagger\eta, \\
&& H_\text{nh}^\dagger=\rho H_\text{nh} \rho^{-1}, \quad \text{with}~\rho=\eta^\dagger\eta. \notag
\end{eqnarray}
Thus, the non-Hermitian Hamiltonian $H_\text{nh}$ becomes Hermitian with respect to a linear, non-unique, invertible and self-adjoint operator $\rho$ playing the role of a metric. Consequently, the eigenstates of the non-Hermitian Hamiltonian $H_\text{nh}$ turns out to be orthogonal with respect to the so-called pseudo-inner product provided that the eigenstates of $H_\text{h}$ and $H_\text{nh}$ ($|\Phi_\text{h}\rangle$ and $|\Phi_\text{nh}\rangle$, respectively) are constrained by the relation $|\Phi_\text{nh}\rangle=\eta^{-1}|\Phi_\text{h}\rangle$. Within the formalism, all crucial properties of a physical system can be proven rigorously \cite{Mostafazadeh_2002,Mostafazadeh_2010,Brody_2014} and, therefore, with respect to a metric $\rho$, a non-Hermitian Hamiltonian can represent a complete physical system. It is worth mentioning that pseudo-Hermiticity does not require to assume the operator $\rho$ to be positive. Nevertheless, it is not so straightforward to obtain an exact expression of the metric, in fact, there are only a limited number of articles producing exact forms, see; for instance \cite{Bagchi_Quesne_Roychoudhury_2005,Znojil_Geyer_2006,Dey_Fring_Mathanaranjan_2014, Dey_Fring_Mathanaranjan_2015}. However, there are various methods to obtain the same, for example, using the perturbation theory, spectral theory
\cite{Bender_Brody_Jones_2004,Ghatak_Mandal_2013}, Moyal product \cite{Faria_Fring_2006}, etc.

Pseudo-Hermiticity is equally important and significant as the $\mathcal{PT}$-symmetric theory. People argue that the scope of $PT$-symmetric theories is limited in comparison to that of pseudo-Hermiticity in the sense that $\mathcal{PT}$-symmetry does not always guarantee a real spectrum. Despite the fact that such issues are debated \cite{Mostafazadeh_2003,Bender_Chen_Milton_2006,Dey_2014_Thesis}, it is conceivable that both of the theories have made a significant development of the field of non-Hermitian physics  \cite{Graefe_Korsch_Niederle_2008,Dey_Fring_Khantoul_2013,Dey_Fring_2013,Dey_Fring_2014,Dey_Fring_Gouba_2015, Zelaya_Dey_etal_2020}.

\subsection{Different methods of generalization}\label{sec32}
In the last few decades, the study on the generalizations of quantum optical states has been one of the foremost priorities of researchers because of their widespread applications in almost every branch of physics. Though the progress is slowed down significantly due to many mathematical hardships, the outcomes are often rewarding. The initial efforts came from the generalization of coherent states in the group theoretical framework \cite{Perelomov_1972}. Later they were extended to the wavelet theories, which have been found to be very useful in many areas including signal analysis and image processing in a joint time-frequency transform. It is out of our scope to discuss such developments in detail, for further interest in this area, one may refer to some existing reviews and books \cite{Perelomov_1986_Book, Zhang_Feng_Gilmore_1990_Review,Ali_Antoine_Gazeau_2014_Book}. There exist several other methods of generalization, however, we shall focus on the nonlinear and $q$-deformed construction of generalized nonclassical states assisted by the non-Hermitian theories; like $\mathcal{PT}$-symmetry and pseudo-Hermiticity. This will lead us to the main goal of this review towards the resource theoretical analysis of the generalized nonclassical light which will be expounded in Sec.\,\ref{sec4}.

\subsubsection{Nonlinear generalization}\label{sec321}
The starting point of the nonlinear generalization is to gain insights on the generalizations of the ladder operators $\hat{A}^\dagger\equiv f(\hat{n})\hat{a}^\dagger=\hat{a}^\dagger f(\hat{n}+1),~\hat{A}\equiv \hat{a}f(\hat{n})=f(\hat{n}+1)\hat{a}$, obeying the following nonlinear commutator algebras
\begin{eqnarray}\label{NonlinComm}
&& [\hat{A},\hat{A}^\dagger] = (\hat{n}+1)f^2(\hat{n}+1)-\hat{n}f^2(\hat{n}), \\
&& [\hat{n},\hat{A}]=-\hat{A}, \quad [\hat{n},\hat{A}^\dagger]=\hat{A}^\dagger. \notag
\end{eqnarray}
Usually, $f(\hat{n})$ is a general operator-valued nonlinear function of the standard number operator $\hat{n}=\hat{a}^\dagger \hat{a}$. The algebras \eqref{NonlinComm} naturally reduce to the Heisenberg algebra, $[\hat{a},\hat{a}^\dagger]=1,[\hat{n},\hat{a}]=-\hat{a},[\hat{n},\hat{a}^\dagger]=\hat{a}^\dagger$, when $f(\hat{n})$ is chosen to be $1$. The aim is to find out the function $f(\hat{n})$ corresponding to a general physical system so that the action of the generalized operators $\hat{A}$ and $\hat{A}^\dagger$ on the Fock states can be realized as
\begin{eqnarray}\label{GenLad}
&& \hat{A}^\dagger|n\rangle = \sqrt{n+1}f(n+1)|n+1\rangle, \\
&& \hat{A}|n\rangle = \sqrt{n}f(n)|n-1\rangle. \notag 
\end{eqnarray}
If we assume that the Hamiltonian of the general quantum mechanical model can be factorized in terms of the generalized number operator $\hat{A}^\dagger \hat{A}$, the exact form of $f(\hat{n})$ can be obtained from the eigenvalues of the associated physical system. The appearance of any constant terms in the Hamiltonian can always be eliminated by simply rescaling the eigenvalues accordingly. Therefore, the first requirement is to solve the eigenvalues of the general model which can be utilized to build the ladder operators of the corresponding model. Most of the quantum optical states (coherent and nonclassical) arise from different quantum optical relations composed of the ladder operators. Thus, on many occasions, direct use of the ladder operators $\hat{A},\hat{A}^\dagger$ produces the generalized quantum optical states, however, it is not always the case. We shall explain such scenarios while discussing some particular models in Sec.\,\ref{sec4}. Also, it is worth mentioning that in order to claim a well-defined quantum optical state, one has to study all the necessary properties of the states individually. For example, the construction of coherent states requires the resolution of identity; see, for instance \cite{Dey_2017}, for further details. Nevertheless, the method of generalization emerging from a nonlinear function $f(n)$ consisting of the creation and annihilation operators is widely known as the nonlinear generalization \cite{Filho_Vogel_1996,Manko_etal_1997,Mancini_1997,Sivakumar_1999} and, nowadays, the endeavor is widely established; see, for example \cite{Trifonov_2000,Kwek_Kiang_2003,Dey_Hussin_2015, Hertz_Dey_etal_2016,Dey_Fring_Hussin_2017}. The formulation has given rise to some similar type of generalizations; such as the Gazeau-Klauder generalization \cite{Gazeau_Klauder_1999,Dey_Fring_2012,Dey_etal_2013}, generalization to supersymmetric quantum mechanics \cite{Benedict_Molnar_1999,Fernandez_Hussin_Rosas-Ortiz_2007}, etc.

\subsubsection{Generalization from $q$-deformation}\label{sec322}
The sole purpose of $q$-deformed generalization is to provide a systematic framework for the construction of quantum optics for various deformed quantum mechanical models. $q$-deformation is a unified framework that can be related with the quantum groups and, thus, a deformed model can always be associated with a deformed canonical commutation relation of type
\begin{equation}\label{DefCano}
\hat{A}_q\hat{A}_q^\dagger-q^2\hat{A}_q^\dagger \hat{A}_q=q^{g(\hat{n})}, \quad |q|<1,
\end{equation} 
with $g(\hat{n})$ being an arbitrary function of the number operator. The ladder operators $\hat{A}_q$ and $\hat{A}_q^\dagger$ are constructed in a $q$-deformed Fock space
\begin{eqnarray}\label{qFock}
&& \vert n\rangle_q := \frac{\hat{A}_q^{\dagger n}}{\sqrt{[n]_q!}}\vert 0\rangle_q, \quad [n]_q!:= \displaystyle\prod_{k=1}^{n}[k]_q, \\
&& [0]_q! :=1, \quad \hat{A}_q\vert 0\rangle_q = 0, \quad ~_{q}\langle 0 \vert 0\rangle_q = 1, \notag
\end{eqnarray}
such that the following operations are satisfied
\begin{eqnarray}\label{qLadder}
\hat{A}_q\vert n \rangle_q &=& \sqrt{[n]_q}~\vert n-1\rangle_q, \\
\hat{A}_q^\dagger\vert n \rangle_q &=& \sqrt{[n+1]_q}~\vert n+1\rangle_q. \notag
\end{eqnarray}
Here, the deformed number operators $[\hat{n}]_q\equiv\hat{A}_q^\dagger\hat{A}_q$ are explicitly obtained as a solution of \eqref{DefCano} with the help of  \eqref{qFock} and \eqref{qLadder}. This implies that one obtains a well-defined quantum mechanical framework in the sense that the states $\vert n\rangle_q$ form an orthonormal basis in a $q$-deformed Hilbert space $\mathcal{H}_q$ spanned by the vectors $\vert\psi\rangle_q :=\sum_{n=0}^{\infty}c_n\vert n\rangle_q$ with $c_n\in\mathbb{C}$, such that ${}_q\langle\psi\vert\psi\rangle_{q} = \sum_{n=0}^{\infty}\vert c_n\vert^2<\infty$. The commutation relation between $\hat{A}_q$ and $\hat{A}_q^\dagger$ takes the form
\begin{equation}
[\hat{A}_q,\hat{A}_q^\dagger]=1+(q^2-1)\hat{A}_q^\dagger \hat{A}_q=1+(q^2-1)[\hat{n}]_q,
\end{equation}
which reduces to the canonical commutation relation in the limit $q\rightarrow 1$. The $q$-deformed oscillator algebras of type \eqref{DefCano} have been shown to be correlated with various physical systems \cite{Arik_Coon_1976,Biedenharn_1989,Macfarlane_1989,Sun_Fu_1989,Kulish_Damaskinsky_1990} and, thus, they have been utilized for the generalizations of various quantum optical states \cite{Quesne_Penson_Tkachuk_2003,Naderi_Soltanolkotabi_Roknizadeh_2004,Dey_2015,Dey_Hussin_2016,Fakhri_Hashemi_2016, Jayakrishnan_Dey_etal_2017}.

Note that, both the nonlinear and $q$-deformed generalizations can be applied to the non-Hermitian Hamiltonian systems also.  It is obvious that the non-Hermitian generalization to quantum optics creates additional complications, however, they can be resolved using the formalism discussed in Sec.\,\ref{sec31}. For further information in this regard, one may refer to a review article dedicated to this scenario \cite{Dey_Fring_Hussin_2018_Review}.  

\section{\textbf{Resources theories of nonclassicality with generalized quantum optical states}}\label{sec4}
Here, we shall discuss some important generalized quantum optical states with a focus on their resources of nonclassicality.

\subsection{Generalized nonclassical states}\label{sec41}

\subsubsection{Squeezed states}\label{sec411}
Squeezed states are known to be one of the most useful nonclassical states. The optical noise produced by a squeezed state is less than a coherent state \cite{Walls_1983,Loudon_Knight_1987} and, this is why squeezed states are used extensively in many areas \cite{Hillary_2000,Menicucci_etal_2006,Riedel_etal_2010,Giovannetti_Lloyd_Maccone_2011} including gravitational wave detection \cite{Vahlbruch_etal_2006,Schnabel_etal_2010,Aasi_etal_2013,Chua_etal_2014}. For further interest in the applications of squeezed states, one may refer to some review articles; see, for instance \cite{Dodonov_2002_Review,Braunstein_van-Loock_2005_Review,Dey_Fring_Hussin_2018_Review}. The generalization of squeezed states starts with the replacement of the harmonic oscillator ladder operators $\hat{a},\hat{a}^\dagger$ with the generalized ones $\hat{A},\hat{A}^\dagger$ in any of the following two definitions \cite{Dey_Hussin_2015,Zelaya_Dey_Hussin_2018}
\begin{eqnarray}
&& (\hat{a}+\xi\hat{a}^\dagger)|\alpha,\xi\rangle =\alpha |\alpha,\xi\rangle, \label{Squeezed1}  \\
&& |\alpha,\xi\rangle=\hat{D}(\alpha)\hat{S}(\xi)|0\rangle, \quad \alpha,\xi\in\mathbb{C}. \label{Squeezed2}
\end{eqnarray}
Here, $\hat{D}(\alpha)=\exp{(\alpha\hat{a}^\dagger+\alpha^\ast\hat{a})}$ and $\hat{S}(\xi)=\exp{\{(\xi\hat{a}^{\dagger 2}-\xi^\ast\hat{a}^2)/2\}}$ are the displacement and squeezing operator, respectively. By following the definition \eqref{Squeezed1}, the construction of the generalized squeezed state in Fock space follows from the assumption that it can be expanded in the Fock basis as
\begin{equation}\label{Expansion}
\vert \alpha,\xi\rangle_\text{gen}=\frac{1}{\mathcal{N}(\alpha,\xi)}\displaystyle\sum_{n=0}^\infty \frac{\mathcal{I}(\alpha,\xi,n)}{\sqrt{n!}f(n)!}\vert n \rangle~,
\end{equation}
and then solving the unknown function $\mathcal{I}(\alpha,\xi,n)$ from the corresponding definition of the squeezed states \eqref{Squeezed1}. By replacing \eqref{Expansion} in \eqref{Squeezed1} along with the help of \eqref{GenLad}, one ends up with a recurrence relation
\begin{equation}\label{recurrence}
\mathcal{I}(\alpha,\xi,n+1)=\alpha~\mathcal{I}(\alpha,\xi,n)-\xi nf^2(n) \mathcal{I}(\alpha,\xi,n-1),
\end{equation}
whose solution will constitute the required state $\vert\alpha,\xi\rangle_\text{gen}$. However, the difficulty in finding a general solution of \eqref{recurrence} for an arbitrary function $f(n)$ has barred the community for a long time to find an exact expression of the generalized squeezed states $\vert\alpha,\xi\rangle_\text{gen}$. In some of the recent studies, the squeezed states for particular quantum systems having a particular form of the function $f(n)$ have been constructed \cite{Dey_Hussin_2015}, however, a general solution of the squeezed state has come out recently in \cite{Zelaya_Dey_Hussin_2018}. The usage of non-Hermitian theories becomes inevitable since, on many occasions, the Hamiltonians turn out to be non-Hermitian; see, for instance \cite{Dey_Hussin_2015,Dey_Fring_Hussin_2018_Review}, for more details. Nevertheless, the computation of the general solution of \eqref{recurrence} involves certain mathematical challenges, which is, however, resolved in \cite{Zelaya_Dey_Hussin_2018} and the solution is given by
\begin{eqnarray}\label{GenSqState}
\mathcal{I}(\alpha,\xi,n)=\displaystyle\sum_{m=0}^{[n/2]}(-\xi)^m\alpha^{n-2m}g(n,m),
\end{eqnarray} 
with
\begin{equation}\label{gnm}
g(n,m)=\displaystyle\sum_{j_1=1}^{(n-2m+1)}\displaystyle\sum_{j_2=j_1+2}^{(n-2m+3)}\displaystyle\sum_{j_3=j_2+2}^{(n-2m+5)}\cdot\cdot\cdot\displaystyle\sum_{j_m=j_{m-1}+2}^{(n-1)}\mu.
\end{equation}
Here, $\mu=k(j_1)k(j_2)\cdot\cdot\cdot k(j_m),~g(n,0)=1$ and $k(j):=jf^2(j)$. The symbol $[n]$ in \eqref{GenSqState} denotes the Floor function, which results in an integer value less than or equal to the corresponding real number $n$. It is now obvious that the generalized squeezed state comprised of \eqref{Expansion} and \eqref{GenSqState} can be applied to any quantum mechanical system to find the corresponding squeezed state. In \cite{Zelaya_Dey_Hussin_2018}, the result has been analyzed with an example of a Rosen-Morse potential where it was shown that the general solution \eqref{GenSqState} can be utilized without any further difficulties.

\subsubsection{Schr\"odinger cat states}\label{sec412}
Generalization of cat states is straightforward once the same is realized for coherent states. The generalized coherent states in the $q$-deformed formulation arises from the definition $\hat{A}_q|\alpha\rangle_q^\text{gen}=\alpha |\alpha\rangle_q^\text{gen}$ with the use of \eqref{qLadder} as follows
\begin{equation}\label{qdefcoherent}
\big\vert\alpha\big\rangle_q^\text{gen}=\frac{1}{\mathcal{N}_q(\alpha,q)}\displaystyle\sum_{n=0}^{\infty}\frac{\alpha^n}{\sqrt{[n]_q!}}\vert n\rangle_q \quad \alpha\in\mathbb{C}.
\end{equation}
The Schr\"odinger cat states are formed as even and odd superposition of coherent states \cite{Dey_2015}
\begin{equation}\label{qCat}
\big\vert\alpha\big\rangle_{q,\pm}^\text{gen} = \frac{1}{\mathcal{N}_{q,\pm}(\alpha,q)}\Big(\vert\alpha\rangle_q^\text{gen}\pm\vert-\alpha\rangle_q^\text{gen}\Big),
\end{equation} 
with the normalization constant being
\begin{equation}
\mathcal{N}_{q,\pm}^2(\alpha,q) = 2\pm\frac{2}{\mathcal{N}_{q}^2(\alpha,q)}\displaystyle\sum_{n=0}^{\infty}\frac{(-1)^n\vert\alpha\vert^{2n}}{[n]_q!}.
\end{equation}
For further details on the construction, we refer the readers to \cite{Dey_2015}. 

\subsubsection{Photon-added coherent states}\label{sec413}
Photon-added coherent states are yet another interesting class of nonclassical states, the $q$-deformed generalization of which can be performed by operating the deformed photon creation operator $\hat{A}_q^\dagger$ successively on the deformed coherent states 
\begin{eqnarray}
&& \vert\alpha,m\rangle_q^\text{gen} = \frac{1}{\mathcal{N}_q(\alpha,q,m)}\hat{A}_q^{\dagger m}\vert\alpha\rangle_q^\text{gen} \label{qPACS} \\
&& \quad= \frac{1}{\mathcal{N}_q(\alpha,q,m)\mathcal{N}_q(\alpha,q)}\displaystyle\sum_{n=0}^{\infty}\frac{\alpha^n}{[n]_q!}\sqrt{[n+m]_q!}~\vert n+m\rangle_q, \notag
\end{eqnarray}
with the normalization constant being
\begin{eqnarray}
\mathcal{N}_q^2(\alpha,q,m) &=& {}^\text{gen}{}_q\langle\alpha,m\vert \hat{A}_q^m\hat{A}_q^{\dagger m}\vert\alpha,m\rangle_q^\text{gen} \\
&=& \frac{1}{\mathcal{N}_q^2(\alpha,q)}\displaystyle\sum_{n=0}^{\infty}\frac{\vert\alpha\vert^{2n}}{[n]_q!^2}[n+m]_q!. \notag
\end{eqnarray}
A more detailed analysis of the state can be found in \cite{Dey_Hussin_2016}.

\subsubsection{Photon-subtracted squeezed vacuum states}\label{sec414}
Squeezed vacuum states (SVS) are nonclassical. Photon subtraction from the SVS makes the state even more nonclassical. The harmonic oscillator SVS $|\xi\rangle$ are built from the solution of the following equation 
\begin{equation}\label{SVS}
(\hat{a}\cosh r+\hat{a}^\dagger e^{i\theta}\sinh r)|\xi\rangle =0,
\end{equation} 
which originally follows the definition of SVS, $|\xi\rangle=\hat{S}(\xi)|0\rangle$. The generalization of SVS does not arise from the direct replacement of $\hat{a},\hat{a}^\dagger$ with $\hat{A},\hat{A}^\dagger$ because the generalized squeezing operator $\hat{S}_\text{gen}(\xi)=\exp{\{(\xi\hat{A}^{\dagger 2}-\xi^\ast\hat{A}^2)/2\}}$ is no longer possible to disentangle and, therefore, the relations $(\hat{A}\cosh r+\hat{A}^\dagger e^{i\theta}\sinh r)|\xi\rangle_\text{gen}=0$ and $|\xi\rangle_\text{gen}=\hat{S}_\text{gen}(\xi)|0\rangle$ are no longer equivalent to each other. The problem can be overcome by introducing a set of generalized auxiliary ladder operators $\hat{B}=\hat{a}[1/f(\hat{n})]=[1/f(\hat{n}+1)]\hat{a}$ and $\hat{B}^\dagger=[1/f(\hat{n})]\hat{a}^\dagger=\hat{a}^\dagger [1/f(\hat{n}+1)]$ giving rise to a new set of commutation relations $[\hat{A},\hat{B}^\dagger]=[\hat{B},\hat{A}^\dagger]=1$. Therefore, the new set of conjugate ladder operators $\hat{A},\hat{B}^\dagger$ or $\hat{B},\hat{A}^\dagger$ will permit the disentanglement of the squeezing operator \cite{Dey_Nair_2020}. The definition of generalized SVS is, thus, altered as
\begin{equation}\label{GSVS2}
(\hat{A}\cosh r+\hat{B}^\dagger e^{i\theta}\sinh r)|\xi\rangle_\text{gen}=0,
\end{equation}
leading to the recurrence relation
\begin{equation}\label{recursionSVS}
\mathcal{J}_{m+1}=-\frac{e^{i\theta}\tanh r}{f(m)f(m+1)}\sqrt{\frac{m}{m+1}}\mathcal{J}_{m-1},
\end{equation}
which when solved in an even basis, the required state comes out as follows
\begin{equation}\label{GSVS}
|\xi\rangle_\text{gen}=\frac{1}{\mathcal{N}(r)}\displaystyle\sum_{n=0}^{\infty}(-1)^n\frac{e^{in\theta}(\tanh r)^n\sqrt{(2n)!}}{2^nn!f(2n)!}|2n\rangle.
\end{equation}
The odd basis solution of \eqref{recursionSVS} gives rise to the generalized squeezed first excited state that may also be interesting, however, it is not our focus here. Photon-subtracted SVS $|\xi,m\rangle_\text{gen}$ are then prepared by removing photons from the SVS, as given by \cite{Dey_Nair_2020}
\begin{eqnarray} \label{EPSSVS}
&& |\xi,m\rangle_\text{gen}^\text{e} = \hat{A}^{2m}|\xi\rangle_\text{gen} \\
&& = \frac{1}{\mathcal{N}^\text{e}(r,m)}\displaystyle\sum_{n=0}^{\infty}\frac{(-\tanh r)^{k}e^{ik\theta}(2k)!}{2^{k}k!\sqrt{(2n)!}f(2n)!}|2n\rangle, \notag 
\end{eqnarray}  
and
\begin{eqnarray} \label{OPSSVS}
&& |\xi,m\rangle_\text{gen}^\text{o} = \hat{A}^{2m+1}|\xi\rangle_\text{gen} \\
&&  = \frac{1}{\mathcal{N}^\text{o}(r,m)}\displaystyle\sum_{n=0}^{\infty}\frac{(-\tanh r)^{k'}e^{ik'\theta}(2k')!}{2^{k'}k'!\sqrt{(2n+1)!}f(2n+1)!}|2n+1\rangle, \notag
\end{eqnarray}
in even and odd basis, respectively. Here, $k=m+n$ and $k'=k+1$.

\subsection{Resource theories}\label{sec42}
The sole intention of building all of the above generalized nonclassical states is to study their usefulness as resources. The three resource theories discussed in subsections \ref{sec22}, \ref{sec23} and \ref{sec24} have been analyzed for all of the four generalized states along with their harmonic oscillator counterparts in \cite{Dey_2015,Dey_Hussin_2016,Zelaya_Dey_Hussin_2018,Dey_Nair_2020}. Quadrature and number squeezing are relatively straightforward to compute, but due to mathematical complexity, sometimes the computations of quasi-probability distribution functions become troublesome for generalized states. However, as discussed earlier in subsection \ref{sec22} that the phenomenon of quadrature squeezing is equivalent to the negativity of the $P$-function, which is the strongest measure of nonclassicality among the three quasi-probability distribution functions. Therefore, even if someone is not able to compute the quasi-probability distribution functions in some cases, the results will not be weakened provided that one is able to compute the quadrature squeezing. In the cases of squeezed states \cite{Zelaya_Dey_Hussin_2018} and photon-subtracted squeezed vacuum states \cite{Dey_Nair_2020}, all of the three resource theories have been discussed in this article, whereas in the case of cat states \cite{Dey_2015} and photon-added coherent states \cite{Dey_Hussin_2016}, only two of them, namely the quadrature and number squeezing, have been analyzed. In fact, for photon-added coherent states \cite{Dey_Hussin_2016} higher order squeezing effects have been studied which are known to be a more robust measure of nonclassicality compared to the first order squeezing effects. 

It is known that most of the harmonic oscillator nonclassical states are good resources of nonclassicality in the sense that they exhibit non-negativity of the quasi-probability distributions, quadrature squeezing and number squeezing with respect to the measures \eqref{QuasiMeasure}, \eqref{QSMeasure} and \eqref{NSMeasure}, respectively. However, the novelty of the results in \cite{Dey_2015,Dey_Hussin_2016,Zelaya_Dey_Hussin_2018,Dey_Nair_2020} lies in the fact that the generalized nonclassical states behave as a richer source of nonclassicality in comparison to those of the harmonic oscillator with respect to the same measures of nonclassicality given in \eqref{QuasiMeasure}, \eqref{QSMeasure} and \eqref{NSMeasure}. Moreover, one of the striking features of the $q$-deformed generalizations is that one obtains an additional degree of freedom in such systems with which the degree of nonclassicality can be enhanced further by controlling the deformation parameter. It was shown that in some cases, the nonclassicality measure attains a value very close to the maximal possible value, which has never been possible earlier for harmonic oscillator nonclassical states. Overall, it is indicative that the generalized nonclassical states may be utilized as enriched resources of nonclassicality and, thus, they legitimately claim further attention and exploration. 

\section{\textbf{Realization of generalized nonclassical light in real life}}\label{sec5}
Real-life realization of squeezed states and other nonclassical states of the harmonic oscillator is a long-term goal of researchers. The aspiration has started in the late twentieth century and, nowadays, they are fairly well-realized via numerous sophisticated methods, like; wave mixing in optical fibers \cite{Reid_Walls_1985,Slusher_etal_1986}, parametric down-conversion \cite{Wu_etal_1986}, normal modes of the coupled atom-field system \cite{Raizen_etal_1987}, optical \cite{Dodonov_Klimov_Manko_1990,Bose_Jacobs_Knight_1997} and nanomechanical resonators \cite{Rabl_Shnirman_Zoller_2004}, microwave radiation in superconducting resonant circuit \cite{Zagoskin_etal_2008}, homodyne detection technique \cite{Ourjoumtsev_etal_2007,Sychev_etal_2017}, optomechanical coupling \cite{Gu_Li_Yang_2013,Rashid_etal_2016}, many-body systems \cite{Omran_etal_2019}, superconducting qubit\cite{Song_etal_2019}, etc. Nonlinear light is nothing but a nonlinear dependence in the polarization of the light. Therefore, the simplest realization of nonlinear nonclassical lights can be brought in by sending the harmonic oscillator nonclassical lights through a nonlinear crystal or any other nonlinear mediums. Indeed, such a concept has been utilized in \cite{Sanders_1992} where a Kerr-type nonlinearity has been introduced in one of the arms of a Mach-Zehnder interferometer to analyze the entanglement properties of the output. Though, at that time, the output states were not interpreted as nonlinear quantum light and the states emerging from the setup may not be identical with the generalized nonclassical states that are studied here. However, a similar concept can be considered as the starting point of the journey of nonlinear quantum optical states. Generation of the nonlinear quantum optical states in real life are underway. More sophisticated concepts of nonlinearity have been introduced over the years, however, the primitive idea remains somewhat similar. Today there exist plenty of studies exhibiting Kerr-type nonlinear effects through atomic coherence in multilevel atomic systems \cite{Wang_Goorskey_Xiao_2001}, carbon nanotubes \cite{Gambetta_etal_2006}, optical cavities with interacting photons \cite{Chang_Vuletic_Mikhail_2014}, optomechanical cavities \cite{Lu_etal_2013,Yan_Zhu_Li_2016}, topological nanostructure \cite{Kruk_etal_2019}, nanoimaging in graphene \cite{Jiang_etal_2019}, etc. Several other types of nonlinearities have also been realized primarily via quantum optomechanical interaction \cite{Pikovski_etal_2012,Dey_etal_2017,Khodadi_Nozari_Dey_etal_2018} and their usefulness in practical scenarios has been indicated \cite{Dey_Hussin_2019}.

The realization of generalized nonclassical lights that we have reviewed in this article will surely require further sophistication, however, we believe that the clue to the realization of such states lies in the realization of Kerr and other types of nonlinearities, which are underway. We hope the experimental realization of generalized nonclassical light will be possible in near future.

\section{\textbf{Conclusions}}\label{sec6} 
This review article began with a broader understanding of the operational approach of quantum resource theory followed by a focused discussion on various nonclassical resources of different quantum optical states. We have summarized some of the most common nonclassical measures as resources, such as; quasi-probability distribution functions, quadrature squeezing and sub-Poissonian photon statistics. One of the important objectives of this review is to scrutinize the usefulness of generalized nonlinear and $q$-deformed quantum optical states in different frameworks of quantum optics and information theories, and to quantify their nonclassicality measure within the unified roof of nonclassical resource theories. There are several other interesting observations in Sec.\,\ref{sec4} during the study of $q$-deformed nonclassical states, especially the cat states and the photon-added coherent states. A common feature that is attributed to both of the states is that when the deformation is increased, the corresponding nonclassical states demonstrate an increased degree of nonclassicality. The deformation parameter being the characteristics of the models themselves, it is conceivable that the elevated degree of nonclassicality is an inherent property of the corresponding nonclassical states, which cannot be generated from the standard quantum optical states via quantum operations. Thereby exploring a new possibility of obtaining better resources of nonclassicality that can be utilized more efficiently towards the different purposes of quantum information and computation.

\vspace{0.5cm} \noindent \textbf{Acknowledgments:} The author acknowledges the support of the research grant (DST/INSPIRE/04/2016/ 001391) by DST-INSPIRE, Govt. of India.

\end{document}